\documentclass[aps,prc,twocolumn,superscriptaddress,10pt,fleqn,floatfix]{revtex4}

\usepackage{amsmath}
\usepackage{amssymb}
\usepackage[dvips]{graphicx}
\usepackage{bm}

\newcommand{\bra}{\left\langle A0\right|}
\newcommand{\ket}{\left| A0\right\rangle}

\begin{document}

\title{Extended Theory of Finite Fermi Systems: \\
Application to the collective and non-collective E1 strength in $^{208}$Pb}

\author{V.Tselyaev}
\affiliation{Institut f\"ur Kernphysik, Forschungszentrum J\"ulich, 52425 J\"ulich, Germany}
\affiliation{Institute of Physics S.Petersburg University, Russia}
\author{J.Speth}
\affiliation{Institut f\"ur Kernphysik, Forschungszentrum J\"ulich, 52425 J\"ulich, Germany}
\affiliation{Institute of Nuclear Physics, PAN, PL-31-342 Cracow, Poland}
\author{F.Gr\"ummer}
\affiliation{Institut f\"ur Kernphysik, Forschungszentrum J\"ulich, 52425 J\"ulich, Germany}
\author{S.Krewald}
\affiliation{Institut f\"ur Kernphysik, Forschungszentrum J\"ulich, 52425 J\"ulich, Germany}
\author{A.Avdeenkov}
\affiliation{Institut f\"ur Kernphysik, Forschungszentrum J\"ulich, 52425 J\"ulich, Germany}
\affiliation{Institute of Physics and Power Engineering, 249020 Obninsk, Russia}
\author{E.Litvinova}
\affiliation{Institute of Physics and Power Engineering, 249020 Obninsk, Russia}
\author{G.Tertychny}
\affiliation{Institut f\"ur Kernphysik, Forschungszentrum J\"ulich, 52425 J\"ulich, Germany}
\affiliation{Institute of Physics and Power Engineering, 249020 Obninsk, Russia}

\date{ 4. December 2006}

\begin{abstract}
The \emph{Extended Theory of Finite Fermi Systems} is based on the conventional Landau-Migdal 
theory and includes the coupling to the low-lying phonons in a consistent way. 
The phonons give rise to a fragmentation of the single-particle strength and to a compression of 
the single-particle spectrum. 
Both effects are crucial for a quantitative understanding of nuclear structure properties. 
We demonstrate the effects on the electric dipole states in $^{208}$Pb (which possesses $50\%$ 
more neutrons then protons) where we calculated the low-lying non-collective spectrum as well as 
the high-lying collective resonances. 
Below 8 MeV, where one expects the so called isovector \emph{pygmy} resonances, we also find a 
strong admixture of isoscalar strength that comes from the coupling to the high-lying isoscalar 
electric dipole resonance, which we obtain at about 22 MeV. 
The transition density of this resonance is very similar to the breathing mode, which we also 
calculated. 
We shall show that the \emph{extended theory} is the correct approach for self-consistent 
calculations, where one starts with effective Lagrangians and effective Hamiltonians, 
respectively, if one wishes to describe simultaneously collective and non-collective properties of 
the nuclear spectrum. 
In all cases for which experimental data exist the agreement with the present theory results is 
good.
\end{abstract}

\maketitle

\newpage



\setcounter{equation}{0}

\section{Introduction}
The structure of neutron-rich nuclei is important for many applications in  astrophysics. 
The rapid neutron-capture process, for instance, requires detailed knowledge of the properties of 
nuclei between the valley of stability and the neutron drip-line, such as neutron capture rates or
photon-induced neutron emission cross sections.
Progress in Nuclear Resonance Fluorescence photon scattering experiments has made possible studies 
of energetically low lying strength \cite{palit04,kneisl96,hartmann02,we04} and to distinguish 
between 1$^{-}$, 2$^{+}$ and 1$^{+}$ states \cite{kneisl96,las85}.
Surprisingly, in the same region, appreciable isoscalar electric dipole strength was already 
detected some years ago with $(\alpha, \alpha' \gamma_0 )$ experiments and reported in 
\cite{hvw01}.
New data using the same technique have already been analyzed \cite{Zil06}.

In analogy to the giant dipole resonance which exhausts the major part of the Thomas-Reiche-Kuhn 
sum rule, the low lying strength is called the {\it pygmy resonance}.
Commonly, this strength is interpreted as an oscillation of the neutron halo against the nuclear 
core \cite{ikeda}.
A collective model predicts  the pygmy B(E1) strength to increase with the neutron excess 
\cite{vanisac}.

For the electromagnetic dissociation of light nuclei with a one-neutron halo, Typel and Baur have 
developed a model independent approach that relies on only a few low-energy parameters, e.g. the 
neutron separation energy. 
Coulomb dissociation is found to be dominated by the energetically low-lying electromagnetic 
dipole strength.
While in experimentally known nuclei, the relevant low-energy parameters can be obtained directly 
from the data, an extrapolation to unknown nuclei remains a challenge for many-body theory
\cite{Typel:2004}.

Several microscopic models have been applied to the pygmy resonances.
The self-consistent calculations are based on an effective  Lagrangian or Hamiltonian
which has the advantage that the extrapolation from the known stable
nuclei to the nuclei near the drip line is well-defined.
Theories of this kind are quite successful in reproducing the
nuclear masses and radii, and also the collective excitation modes of
nuclei, such as the giant dipole resonance, using the quasi-particle
random phase approximation. 
A recent review on the relativistic approach is given in Ref.\cite{ring05}.
Quantitatively, the relativistic approaches to the pygmy resonances do not compare too well
with the experimental data. 
The theory obtains mean excitation energies of the pygmy strength consistently above
the experimental strength. 
This appears to be a general phenomenon observed by several groups.
Goriely and Khan \cite{goriely02} calculated  within the QRPA the E1 strength distributions
for all nuclei with $8 \leq Z \leq 110$  between the proton and neutron drip
lines using known Skyrme forces.
In their calculation the low-lying E1 strength was located systematically higher by some 3 MeV
compared with the available data.

Here, we want to point out that the discrepancy between the theoretical and experimental pygmy 
strength may be overcome by pushing the many-body approach beyond the mean field approximation.

In the past twenty years theoretical investigations of the structure of nuclei are performed in 
self-consistent models as mentioned before and in Migdal's 
\emph{theory of finite Fermi system} (TFFS). 
In both cases one ends up with the \emph{random phase approximation}(RPA), which describes the
excited states of the nuclei as superpositions of particle-hole states. 
The input for the RPA equations are single-particle energies, single-particle wave functions and 
the residual particle-hole (ph) interaction. 
The two approaches differ in the way how the input data are determined. 
In the self-consistent approach one starts with an effective Lagrangian or effective
Hamiltonian and calculates the single-particle energies and single-particle wave functions in the
mean field approximation. 
Also, the residual interaction is derived from the original interaction.
This approach is considered the most fundamental one, as the
effective Lagrangians and effective Hamiltonians can be used for
all nuclei.  As most of the self-consistent
calculations employ effective masses well below $1$, the
corresponding single particle energies are  more widely spread
than the experimental separation energies. Therefore the self-consistent
approaches, in their present form, do not reproduce the
experimental separation energies. A precise knowledge of the separation energies
may be very important, however, as shown in Ref.\cite{Typel:2004}.

In TFFS the single-particle energies and wave-functions are the solutions of a phenomenological
single-particle Hamiltonian and the ph-interaction is parameterized universally for all nuclei.
The RPA results, especially the non-collective solutions, depend sensitively on the 
single-particle spectrum. 
For that reason TFFS-results are closer to the experimental data as, in
general, one uses the experimental spectrum as far as it is available. 
The experimental spectrum as input into the RPA is of crucial importance for states that are not
very collective, such as the odd-parity (magnetic) states. 
These states are, in general, dominated by one single ph-configuration and are only little 
shifted in energy compared to the corresponding uncorrelated ph energies. 
Famous examples in this respect are the $12^-$ and $14^-$ states in $^{208}$Pb \cite{Krew80}. 
Therefore, if one intends to do a consistent nuclear structure calculation in which collective and
non-collective states agree with the data, one has to begin with a single-particle spectrum that 
is close to the experimental one.

In order to investigate this problem, we will rely on an
extended version of the Landau-Migdal theory, ETFFS.
Landau's theory shares two important ideas with
the modern effective field theories due to Weinberg:
one has to identify (I) the degrees of freedom relevant for the
energy scale one is interested in
and then can incorporate the physics of the larger energy scales in
a few low-energy constants and (II) a small expansion parameter. However,
a systematic counting scheme to evaluate higher order corrections was not developed.

In Landau's theory, the relevant degrees of freedom are called quasi-particles and the small 
parameter is the ratio \emph{excitation energy / Fermi energy } \cite{GEB71}.
In finite systems Migdal identified Landau's quasi-particles with the single-particle excitations 
in odd mass nuclei and the associated energy scale with the particle-hole gap, i.e. 16 MeV 
(for medium mass nuclei) and 8 MeV (for the heavy ones), whereas the Fermi energy in nuclei is of 
the order of $40$ MeV. 
In the TFFS only these degree of freedom have been considered.
On the other hand, there is additional degree of freedom, the phonons \cite{soloviev,vg92}.
In the ETFFS, the phonon degree is incorporated explicitly.
The quasi-particles are dressed by the phonons, and it is the {\it dressed } quasi particle energy 
which has to be identified with the experimental separation energy.
Details can be found in Ref.\cite{Kam84,VT89,rev} (and references therein).
The \emph{Quasiparticle Phonon Model} (QPM) by Soloviev \cite{soloviev} has been recently applied 
by Lenske et al. \cite{tsoneva04} to the pygmy resonances. 
In contrast to the original model, here the authors treat the mean-field part microscopically by 
incorporating HFB results as input for the QPM calculations.
This model allows to consider higher phonon states. 
The QRPA plus phonon coupling model by Bortignon \cite{sarchi04} is very similar to the 
present one. 
Here the author start with a HF-BCS mean field were they used a Skyrme type interaction. 
In contrast to our approach the single particle continuum in both models is included in a 
discretized way.

In the following we will give a derivation of the basic equations of the TFFS as well as the 
\emph{extended theory} within the many-body Green function method. 
An important issue is the definition of the one-body potential from which one derives the 
single-particle properties, which are the crucial input data for the RPA. 
We will apply both approaches to the electric dipole states in $^{208}Pb$ and compare the results.
We will show explicitly that in ETFFS the low-lying phonons give rise to a compression of 
the ph spectrum up to several MeV. 
This is an important result for self-consistent approaches:
as the number of phonons and their energy are quite different in the various mass regions, one 
has to include these effects explicitly in self-consistent calculations. 
The mean field in self-consistent calculations provides the \emph{bare} single-particle spectrum, 
which enters in the \emph{extended theory}. 
Therefore, in self-consistent calculations, the 1p1h RPA is insufficient. 
One has to include the effects of the low-lying phonons explicitly, as is done in the 
\emph{extended theory}. 
We demonstrate our findings by investigating the low-lying and high-lying electric dipole states 
in the 1p1h RPA as well as in the extended theory.

\section{Method}

The Landau Migdal theory can be derived within the theory of
\emph{many-body Green functions} \cite{Migdal67,rev77,GS06}. 
The one-particle and two-particle Green functions are defined as
\begin{equation}\label{eq:01}
g_{\nu_1\nu_2}\;{(t_1;t_2)} =
(-i)\bra{T{\left\{{a_{\nu_1}{(t_1)}{a_{\nu_2}^+{(t_2)}}}\right\}}}\ket
\end{equation}
\mathindent0mm
\begin{eqnarray}\label{eq:02}
&&g_{\nu_1\nu_3\nu_2\nu_4}\;{(t_1t_3;t_2t_4)}= \nonumber \\
&&(-i)^2\bra{T{\left\{{a_{\nu_1}{(t_1)}{a_{\nu_3}{(t_3)}}
{a_{\nu_4}^+{(t_4)}}{a_{\nu_2}^+{(t_2)}}}\right\}}}\ket
\end{eqnarray}
with the time-dependent creation and annihilation operators
\begin{equation}\label{eq:03}
a_{\nu}(t)= e^{iHt}\;a_{\nu}\;e^{-iHt}\qquad\qquad a_{\nu}^+(t)= e^{iHt}\;a_{\nu}^+\;e^{-iHt}
\end{equation}
The symbol $T$ denotes the time-ordering operator, which
means that the operators should be taken in time-ordered form with
the latest time to the left and the earliest to the right. The
nucleons are in the single-particle state $\phi_\nu$. Here
$\left|A0\right\rangle$ is assumed to be the exact ground state of
an even-even nucleus. In the Landau Migdal theory one investigates
the response function $L$, which is defined as:
\begin{equation}\label{eq:04}
L(13,24)\;=\;g(13,24)\;-\;g(1,2)g(3,4).
\end{equation}

\begin{figure}[htb]
\begin{center}
\includegraphics[bb=45 504 606 560,width=8cm]{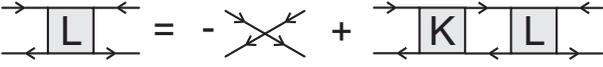}
\end{center}
\caption{\label{fig:1} Graphical representation of the
Bethe-Salpeter equation for the response function in the ph
channel}
\end{figure}

The response function obeys an integral equation of the form
\begin{eqnarray}\label{eq:05}
&&L(13,24)= -g(1,4)g(3,2) \nonumber \\
&&-i\int{d5d6d7d8\;g(1,5)K(57,68)L(83,74)g(6,2)}\,,
\end{eqnarray}
where $K$ is an effective two-body interaction. A graphical
representation of eq.(\ref{eq:05}) is given in Fig.\ref{fig:1}.
One should note that the kernel $K$ of the integral equation is
the irreducible part of the response function with respect to the
particle-hole propagator. We further introduce the vertex function
$\widetilde{\tau}$ by
\begin{equation}\label{eq:06}
L(13,24)= \int{d5d6\;g(1,5)\;\widetilde{\tau}{(53,64)}\;
g(6,2)}\,.
\end{equation}
The corresponding equation follows from eq.(\ref{eq:05})
\begin{eqnarray}\label{eq:07}
&&{\widetilde{\tau}{(13,24)}}=-\delta (1,4)\delta (3,2) \nonumber \\
&&-i\int{d5d6d7d8\;K(17,28)g(8,5)g(6,7)\widetilde{\tau}(53,64)}\,.
\end{eqnarray}
In this paper we investigate the excitation energies and transition amplitudes of double
closed-shell nuclei.
With the time order $t_3,t_4>t_1,t_2$ we insert into the two-particle Green function a complete
system of eigenfunctions of the A-particle system between the two particle-hole pairs:
\begin{equation}\label{eq:08}
g(13,24)=\sum_{n=0}^\infty\;g^{0n}(34)g^{n0}(12)\,.
\end{equation}
with
\begin{equation}\label{eq:09}
g^{0n}(34)\;=\;
-i \left\langle A0 \right| T{\left\{ {a(3)a^+(4)}\right\}} {\left| An\right\rangle}
\end{equation}
and
\begin{equation}\label{eq:10}
g^{n0}(12)\;=\; -i \left\langle An \right| T{\left\{
{a(1)a^+(2)}\right\}} {\left| A0\right\rangle}\,.
\end{equation}

The spectral representation of the response function
eq.(\ref{eq:04}) is also given by eq.(\ref{eq:08}), but the sum
begins at $n=1$
\begin{equation}\label{eq:12}
L(13,24)=\sum_{n=1}^\infty\;g^{0n}(34)g^{n0}(12)\,.
\end{equation}
With the appropriate time-ordering and a Fourier transformation
one obtains the spectral representation the response function
\begin{eqnarray}\label{eq:13}
L_{13,24}=\sum_{n=1}^\infty\;\left(\frac{\chi^{n0*}_{12}\chi^{n0}_{43}}{\Omega+E_n-i\delta}
-\frac{\chi^{n0}_{21}\chi^{n0*}_{34}}{\Omega-E_n+i\delta}\right)_{\delta
\rightarrow 0^+}.\,
\end{eqnarray}
Here $E_n$ are the (exact) energies of the excited states of the A-particle system and
\begin{eqnarray}\label{eq:14}
\chi^{n0}_{12} = < An|a^{+}_{1} a_{ 2}|A0>
\end{eqnarray}
are the corresponding transition matrix elements between the ground state and the excited states
of the A-particle system.
In the actual calculation we solve the equation for the change of the density in an external field
and determine its poles.

\subsection{One-particle Green function and the nuclear shell model}

In order to solve the equation for the response function and
vertex function, respectively, we have to know the one-particle
Green functions $g(1,2)$ and the effective two-body interaction
$K$. As Landau has shown, one does not need the full information
included in $g$, but only the so-called \emph{pole part} of the
one-body Green functions near the Fermi surface. The rest gives
rise to a renormalized interaction and effective one-body
operators. We first discuss the pole part of the one-body Green
function. For that we consider the Dyson equation in coordinate
space \cite{Migdal67}
\begin{eqnarray}\label{eq:15}
&&\left({\epsilon -\frac{p^2}{2m}}\right)g{\left({\xi_1,\xi_2,\epsilon}\right)} \nonumber \\
&&-\int{d
\xi_3\;{\Sigma(\xi_1,\xi_3,\epsilon)}}\;g(\xi_3,\xi_2,\epsilon)=\delta
({\xi_1}-\xi_2)\,,
\end{eqnarray}
where $\xi\equiv{(\bm{r},s)}$ represents space and spin coordinates.
In an arbitrary single-particle basis $\widetilde {\varphi}_{\nu}(\xi)$ the Dyson equation has
the form
\begin{equation}\label{eq:16}
\epsilon \;g_{\nu_1\nu_2}-\sum_{\nu_3}\left[\frac{p^2}{2m}
+\Sigma(\xi_1,\xi_3,\epsilon)\right]_{\nu_1\nu_3}g_{\nu_3\nu_2}=\delta_{\nu_1\nu_2}\,.
\end{equation}
We now chose a special basis $\varphi_{\nu}(\xi)$ that diagonalizes the expression in the brackets
\begin{equation}\label{eq:17}
\left[\frac{p^2}{2m}
+\Sigma(\xi_1,\xi_3,\epsilon)\right]_{\nu_1\nu_3}=
E_{\nu_1}(\epsilon)\delta_{\nu_1\nu_3}\,.
\end{equation}
It is obvious that such a basis must depend on the energy
$\epsilon$. The one-body Green function becomes diagonal in this
basis and can be written as
\begin{equation}\label{eq:18}
g_{{\nu_1}\nu_2}=\frac{{\delta}_{\nu _1 \nu_2}}{\epsilon -
E_{\nu_1 }(\epsilon)}\,.
\end{equation}
Due to Landau's renormalization procedure one needs the
one-particle Green functions only near the dominant poles
$\epsilon_\nu$, i.e. near the poles, which are the solutions of
the equation
\begin{equation}\label{eq:19}
\epsilon_\nu =E_{\nu}(\epsilon_\nu)
\end{equation}
and having maximal single-particle strengths (see below). The
single-particle energies $\epsilon_\nu$ so defined are the
quasi-particle energies (in the sense of Landau) for finite
systems. In the vicinity of $\epsilon_\nu$ we obtain for the
one-(quasi)particle Green function
\begin{equation}\label{eq:22}
g_{\nu_1 \nu_2}(\epsilon)
= \frac{ \delta _{\nu _1 \nu_2}}{ \epsilon - \epsilon_{\nu_1} +i\gamma_{\nu_1}}\;
 {\frac{1}{\left(1-{\frac{dE_\lambda}{d\epsilon}}\right)}}_{\epsilon=\epsilon_{\nu_1}}\,.
\end{equation}
The residue of $g$ at the pole is called the \textit{single-particle strength}
\begin{equation}\label{eq:23}
z_\nu
={\frac{1}{\left(1-{\frac{dE_\nu}{d\epsilon}}\right)}}_{\epsilon=\epsilon_{\nu_1}}\,.
\end{equation}

Within the framework of this method it is easy to derive an
equation for the single-particle energies and the single-particle wave functions.  To
this aim let us omit, for the sake of simplicity, the spin
variables and introduce the self-energy in a \emph{mixed}
representation:
\begin{equation}\label{rsrr}
\Sigma(\bm{R},\bm{k},\epsilon) = \int d\bm{s}\,
\Sigma(\bm{R}-{\textstyle\frac{1}{2}}\bm{s},
\bm{R}+{\textstyle\frac{1}{2}}\bm{s},\epsilon)\,
e^{i\bm{s}\bm{k}}\,,
\end{equation}
where $\bm{R} = {\textstyle\frac{1}{2}}(\bm{r}_1 + \bm{r}_2)$,
$\bm{s} = \bm{r}_2 - \bm{r}_1$. The local energy approximation
(see \cite{PS64}) consists in the replacement of the function
$\Sigma(\bm{R},\bm{k},\epsilon)$ by the following ansatz:
\begin{eqnarray}\label{locen}
&&\Sigma(\bm{R},\bm{k},\epsilon) \simeq
\bar{\Sigma}(\bm{R},k^2_F,\epsilon_F)
\nonumber\\
&&+\,
(k^2 - k^2_F) \left(\partial \bar{\Sigma}(\bm{R},k^2,\epsilon_F)/
\partial k^2 \right)_{k^2 = k^2_F}
\nonumber\\
&&+\,
(\epsilon - \epsilon_F) \left(\partial \bar{\Sigma}(\bm{R},k^2_F,\epsilon)/
\partial \epsilon \right)_{\epsilon = \epsilon_F}\,,
\end{eqnarray}
where $\bar{\Sigma}(\bm{R},k^2,\epsilon)$ is the average with
respect to the angular variables of vector $\bm{k}$,
\begin{equation}\label{barsig}
\bar{\Sigma}(\bm{R},k^2,\epsilon) = \int d\bm{s}\,
\Sigma(\bm{R}-{\textstyle\frac{1}{2}}\bm{s},
\bm{R}+{\textstyle\frac{1}{2}}\bm{s},\epsilon)\,
\frac{\sin (ks)}{ks}
\end{equation}
and the Fermi energy $\epsilon_F$ is related to the Fermi momentum
$k_F(\bm{r})$ by the formula
\begin{equation}\label{eq:25}
\frac{\hbar^2 }{2m}\,k^{2}_F (\bm{r}) +
\bar{\Sigma}(\bm{r},k^2_{F}(\bm{r}),\epsilon_F) = \epsilon_F\,.
\end{equation}
On substituting eq.(\ref{locen}) into the Dyson equation
(\ref{eq:15}) we come, after a series of transformations, to the
following equation for the eigenfunctions:
\begin{equation}\label{smeq}
\left[ - \bm{\nabla} \frac{\hbar^2 }{2m^*(\bm{r})} \bm{\nabla}
+ U_c(\bm{r})\right]\,\varphi_{\lambda}(\bm{r}) =
\epsilon_{\lambda}\,\varphi_{\lambda}(\bm{r})\,,
\end{equation}
where
\begin{equation}\label{effmas}
\frac{m}{m^*(\bm{r})} = a(\bm{r})\left[ 1+\frac{2m}{\hbar^2 }
\left(\frac{\partial \bar{\Sigma}(\bm{r},k^2,\epsilon_F)}
{\partial k^2} \right)_{k^2 = k^2_F(\bm{r})} \right]\,
\end{equation}
and
\begin{equation}\label{sps}
a(\bm{r}) = \left[1 - \left(\frac{\partial
\bar{\Sigma}(\bm{r},k^2_F(\bm{r}),\epsilon)} {\partial \epsilon}
\right)_{\epsilon = \epsilon_F} \right]^{-1}\,.
\end{equation}
If we also take the spin degree of freedom into account then,
proceeding in the same way, one obtains the following
single-particle Hamiltonian:
\begin{eqnarray}\label{eq:30}
&&H = - \bm{\nabla} \frac{\hbar^2 }{2m^*(\bm{r})} \bm{\nabla}
+ U_c(\bm{r})
\nonumber\\
&&+ {\textstyle\frac{1}{2}}\,[\alpha(\bm{r})\,(\bm{\sigma}\bm{l})
+ (\bm{\sigma}\bm{l})\,\alpha(\bm{r})]\,.
\end{eqnarray}
Here $\bm{l}$ is the angular momentum operator. The form of this Hamiltonian follows from the expansion of
the self-energy $\Sigma$ about the Fermi energy $\epsilon_F$ and
the Fermi momentum $p_F$, and symmetry arguments. This Hamiltonian
agrees with phenomenological shell model Hamiltonians. It is
useful to point out \cite{Migdal67} that the spin-orbit term
follows here from the expansion of the non-locality of $\Sigma$. A
certain fraction of the non-locality may be connected with the
$\omega$ and $\sigma$ exchange of the bare nucleon-nucleon
interaction but certainly not the whole effect, as is assumed in
some of the relativistic effective mean field approaches.

\subsection{Landau's renormalization procedure}

We now return to the one-particle Green function. With eqs.(\ref{eq:22} and \ref{eq:23}) we can 
separate the one-particle Green function into a singular part and a remainder:
\begin{equation}\label{eq:31}
g_{\nu_{1}\nu_{2}}(\omega)=
\frac{\delta_{\nu_{1}\nu_{2}}z_{\nu_1}}{\omega - \epsilon_{\nu_1}
+i \eta \; sign( \epsilon_{\nu_1}- \mu)} \;
+\;g^{(r)}_{\nu_{1}\nu_{2}}(\omega)\,,
\end{equation}
with
\begin{eqnarray}\label{eq:32}\nonumber
&&z_{\nu_1}={\left|\left\langle A0\left|a_{\nu_1}\right|A+1\;\nu{_1}\ \right\rangle\right|}^2 \\
&&\epsilon_{\nu_1}= E^{A+1}_{\nu _1}-E^{A}_0  \;\;\;\; for \;\;\epsilon_{\nu_1}> \mu \\ \nonumber
&&z_{\nu_1}={\left|\left\langle A0\left|a^+_{\nu_1}\right|A-1\;\nu{_1}\ \right\rangle\right|}^2 \\
&&\epsilon_{\nu_1}= E^{A}_0-E^{A-1}_{\nu _1}  \;\;\;\; for
\;\;\epsilon_{\nu_1}< \mu\,,
\end{eqnarray}
where $ \left|A\pm1\;\nu{_1} \right\rangle $ denotes an excited state of the
$A\pm1$-particle system. With this \textit{ansatz} one writes the
product of two one-particle Green functions---a quantity that
appears in all the integral equations we have derived before---as
a singular part $S$ and the remainder $R$, using the
\textit{ansatz} in eq.(\ref{eq:31})
\begin{eqnarray}\label{eq:33}
&&g_{\nu_1 \nu_3}\left(\omega +\frac{1}{2}\Omega\right)
 g_{\nu_2\nu_4}\left(\omega -\frac{1}{2}\Omega\right)= \nonumber \\
&&S_{\nu_1\nu_2 \nu_3 \nu_4}\left( \omega; \Omega\right) +R_{\nu_1
\nu_2 \nu_3 \nu_4}\left( \omega; \Omega\right)\,,
\end{eqnarray}
with
\begin{eqnarray}\label{eq:34}
&&S_{\nu_1 \nu_2 \nu_3 \nu_4}\left( \omega; \Omega\right)= 2\pi i z_ {\nu_3} z_{ \nu_4} \delta_{ \nu_1 \nu_3}
\delta_{ \nu_2 \nu_4} \nonumber \\
&& \times \delta \left( \omega - \frac{\epsilon_{ \nu_3} + \epsilon_{
\nu_4}} {2}\right)\;\frac{n_{\nu_2}-n_{\nu_1}}{\epsilon_{\nu_2}-
\epsilon_{\nu_1}-\Omega}\,.
\end{eqnarray}
Here $n_\nu$ are the quasi-particle occupation numbers ($1$ or $0$)
and $\Omega$ is the energy transfer between particle and hole
states. Using Landau's renormalization procedure, one obtains from
eq.(\ref{eq:08}) an equation for the renormalized vertex function
$\tau$:
\begin{eqnarray}\label{eq:35}
&&\tau_{\nu_1\nu_3, \nu_2 \nu_4 }\left( \omega ,\Omega\right)=\tau^{\omega}_{\nu_1 \nu_3,\nu_2 \nu_4 } \left( \omega , \Omega\right) + \nonumber \\
&&\sum_{\nu_5 \nu_6}F^{ph}_{\nu_1 \nu_5 , \nu_2 \nu_6}\left( \omega,\frac{\epsilon_{\nu_5}+ \epsilon_ {\nu_6}}{2}, \Omega\right) \nonumber \\
&&\times \frac{n_{\nu_5} -n_{\nu_6}}{\epsilon_{\nu_5} - \epsilon_{ \nu_6}
-\Omega } \;\tau_{\nu_5 \nu_3, \nu_6 \nu_4 }\left( \frac{\epsilon_
{\nu_5} + \epsilon_{\nu_6}}{2} ;\Omega\right)\,.
\end{eqnarray}
The renormalized vertex is defined as
\begin{equation}\label{eq:36}
\tau_{\nu_1\nu_3, \nu_2 \nu_4 }\left( \omega
,\Omega\right)=\sqrt{z _{\nu _1} z_{\nu _2}} \; \widetilde{\tau
}_{\nu_1\nu_3, \nu_2 \nu_4 }\left( \omega ,\Omega\right)\,.
\end{equation}
Here only the singular part of the product of the two Green
functions appears explicitly, whereas the remainder gives rise to
a renormalized effective two-body interaction $F^{ph}$ and a
renormalized inhomogeneous term $\tau^\omega$. In similar fashion
one can renormalize the response function $L$ and obtains
\begin{eqnarray}\label{eq:37}
&&L(13,24)=\tau ^{\omega} (13,57) \widetilde{L}(57,68)\tau^{\omega}(68,24) \nonumber \\
&&+\tau ^{\omega} (13,57)R(57,24)\,,
\end{eqnarray}
where $\widetilde{L}$ is the quasi-particle response function
\begin{equation}\label{eq:38}
\widetilde{L}(13,24) = S(1,2) \tau(1324)\,.
\end{equation}

\subsection{The renormalized equations of the TFFS}

As we have seen in section II, the response function includes the
transition amplitudes between the ground state and the excited
states of an A-particle system. Using the arguments in section II
and the projection procedure described in \cite{rev77}, one
obtains from the renormalized response function in eq.(\ref{eq:37}) and the renormalized vertex 
function eq.(\ref{eq:35}) an equation for the quasi-particle quasi-hole transition matrix elements
\begin{equation}\label{eq:39}
\left( \epsilon_{\nu_1}-\epsilon_{\nu_2}-
\Omega\right)\chi^{m}_{\nu_1 \nu_2} =
\left(n_{\nu_1}-n_{\nu_2}\right)\sum_{\nu_3 \nu_4}F^{ph}_{\nu_1
\nu_4 \nu_2 \nu_3} \;\chi^{m}_{\nu_3 \nu_4}\,,
\end{equation}
which are connected with the full ph-transition matrix elements
$\chi^{m0}_{\nu_1 \nu_2}$ defined in eq.(\ref {eq:14})
by the relation
\begin{equation}\label{eq:41}
\chi^{m0}_{\nu_1 \nu_2}= \sum_{\nu_3 \nu_4}\tau^{\omega}_{\nu_3
\nu_1 \nu_4 \nu_2} \;\chi^{m}_{\nu_3 \nu_4}\,.
\end{equation}
The transition matrix element of a one-body operator $Q$ is given by
\begin{equation}\label{eq:42}
\left\langle Am \right|Q \left|A0\right\rangle \;=\sum_{\nu_1 \nu_2}Q^{\rm eff}_{\nu_1 \nu_2}\; \chi
^{m}_{\nu_1 \nu_2}\,,
\end{equation}
where $Q^{\rm eff}$ is the renormalized one-body operator
\begin{equation}\label{eq:43}
Q^{\rm eff}_{\nu_1 \nu_2}=\sum_{\nu_3 \nu_4}\tau^{\omega}_{\nu_1
\nu_3 \nu_2\nu_4}Q_{ \nu_3 \nu_4}\,.
\end{equation}

There exist many derivations of the RPA equations (\ref{eq:39}).
In all cases the equations have the identical form but the
approximations that led to them are quite different. In the
present derivation within the many-body Green functions no
approximations have been made so far. We arrive, however, at a
renormalized ph interaction which, in principle, can be calculated
starting from the bare nucleon-nucleon interaction. This, however,
is not done in practice.  An important result of our derivation
concerns the single-particle energies, which are major input data.
The quasi-particle energies in the single-particle Green functions
(eq.(\ref{eq:31})) are the \emph{experimental} single-particle
energies of the $\rm A\pm1$ particle system. In the case of
self-consistent calculations this means that one has to choose an
interaction that reproduces in mean field approximation the
experimental single-particle spectrum of the neighboring nuclei.
This point will be discussed again in connection with the
\emph{extended theory}.

From eq.(\ref{eq:35}) one obtains the change of the
quasi-particle density in an external field (details are given in
Ref.\cite{Migdal67,rev77}):
\begin{eqnarray}\label{eq:44}
&&{\rho}_{\nu_1 \nu_2}(\Omega)=  \frac{n_{\nu_1} -n_{\nu_2}}{\epsilon_{\nu_1} - \epsilon_ {\nu_2}
-\Omega }Q^{\rm eff}_{\nu_1 \nu_2}(\Omega) \nonumber \\
&&- \frac{n_{\nu_1} -n_{\nu_2}}{\epsilon_{\nu_1} - \epsilon_
{\nu_2} -\Omega }\sum_{\nu_3 \nu_4}F^{ph}_{\nu_1 \nu_3 \nu_2
\nu_4} \rho_{\nu_4 \nu_3}(\Omega)\,.
\end{eqnarray}
The actual calculations presented here have been performed in
$\bf{r}$-space because this allows treatment of the continuum in
the most efficient way. The corresponding equation takes the form
\cite{Ka93,rev}
\begin{eqnarray}\label{eq:45}
\rho(\bm{r},\Omega) = -\int d^{3}\bm{r}' A(\bm{r},\bm{r}',\Omega)
Q^{\rm eff}(\bm{r}',\Omega) \\ \nonumber -\int
d^3\bm{r}'d^3\bm{r}''A(\bm{r},\bm{r}',\Omega)
F^{ph}(\bm{r}',\bm{r}'') \rho(\bm{r}'',\Omega)\,.
\end{eqnarray}
The poles of the equation are the excitation energies of the
A-particle system and $\rho(\bm{r},\Omega)$ at a given pole is the
corresponding transition density.

\section{Extended theory of finite Fermi systems (ETFFS)}

As mentioned earlier, the original TFFS allows one to calculate
only the centroid energies and total transition strength of giant
resonances because the approach is restricted to 1p1h
configurations. In order to describe nuclear structure properties
in more detail one has to include 2p2h, or even more complex
configurations. A theoretical approach that takes into account the
complete 2p2h configuration space and a realistic 2p2h interaction
is numerically intractable if one uses a realistically large
configuration space. For this reason the main approximation in the
ETFFS concerns the selection of the most important 2p2h
configurations. One knows from experiments, e.g. from the
neighboring odd mass nuclei of $^{208}$Pb, that the coupling of
low-lying collective phonons to the single-particle states gives
rise to a fragmentation of the single-particle strength, which is
seen in  even-even nuclei as a spreading width in the giant
resonances.

In the past 15 years an extension of the TFFS that includes, in a
consistent microscopic way, the most collective low-lying phonons
has been developed by some of us \cite{VT89,rev,ka06} (and references therein). 
Many of these ideas have been developed in the early work by Werner \cite{Werner66}
and Werner and Emrich \cite{Werner70} 
(see also Refs.\cite{rev77,KS82}) and Ref.\cite{RW73}. 
In this way one considers a special class of 2p2h configurations because the
phonons are calculated within the conventional (1p1h) TFFS. 
The phonons give rise to a modification of the particle and hole
propagators, the ph interaction and the ground state correlations.

\subsection{Some basic relations}

In Landau's theory the self energy $\Sigma$ is irreducible in the
one-particle and one-hole channels, respectively, and the kernel
$K$ in the integral equation for the response function (see
Fig.\ref{fig:1}) is irreducible in the particle-hole channel.
One now introduces a hierarchy of energy dependencies. As one can
see from eq.(\ref{eq:34}), the particle-hole propagator
introduces a strong dependence on the energy transfer $\Omega$.
Compared to this singular behavior, one neglects in Landau's
theory the energy dependence of the ph-interaction. One neglects
consistently the energy dependence of the self-energy $\Sigma$ in
the Dyson equation (\ref{eq:15}) and considers the
quasi-particle and quasi-hole poles only. This approach is of
leading order in the energy transfer $\Omega$. In the
\emph{extended theory}, in which phonons are introduced, one
considers the next-to-leading order in the energy transfer; i.e.
the self energy and the ph interaction become energy dependent. We
then write, explicitly,
\begin{equation}\label{eq:46}
\widetilde{\Sigma}_{\nu}(\epsilon)= \Sigma_{\nu}
+{\Sigma}^{ext}_{\nu}(\epsilon)\,;
\end{equation}
\begin{equation}\label{eq:47}
\widetilde{F}^{ph}_{\nu_1 \nu_3,\nu_2 \nu_4}(\Omega)=
F^{ph}_{\nu_1 \nu_3,\nu_2 \nu_4}+F^{ph,ext}_{\nu_1 \nu_3,\nu_2
\nu_4}(\Omega)\,.
\end{equation}

\begin{figure}[htb]
\begin{center}
\includegraphics[bb=44 463 795 551,width=8cm]{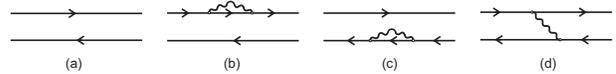}
\end{center}
\caption{\label{fig:2}Graphs corresponding to the propagator of  (a) the RPA and (b-d) the
extended theory.
The wavy lines and the thin lines denote the phonons and the single-particle propagators,
respectively.}
\end{figure}

First we discuss the so-called $g^2$ approximation \cite{Kam84},
which is graphically shown in Fig.\ref{fig:2}. The upper part of
graph (b) gives a correction to the Dyson equation for
quasi-particles, which now has the form
\begin{equation}\label{eq:48}
\left[\epsilon -{\epsilon_{\nu_1}}-\sum_{\nu_2 ;i} \frac{\left|
\gamma^{\nu_2 ;i}_{\nu_1}\right|^2} {\epsilon -\Omega_i
-\epsilon_{\nu_2}} \right]g_{\nu_1}^{ext}(\epsilon )=1\,.
\end{equation}
Here $g_{\nu}^{ext}(\epsilon )$ is the one-quasi-particle Green
function in the \emph{extended theory} and the vertex
$\gamma_{\nu} ^{\mu;i}$, which couples the quasi-particle state
$\nu$ to the core excited configuration $\mu{\otimes}i$, is given
by
\begin{equation}\label{eq:49}
\gamma_{\nu} ^{\mu;i} = \sum_{\alpha,\beta} F^{ph}_{\nu \alpha;
\mu \beta} \chi^{i}_{\alpha\beta}\,,
\end{equation}
where $\chi^i$ is the RPA wave function of the phonon considered.
The corresponding energy-dependent correction for the ph-interaction has the form
\begin{eqnarray}\label{eq:50}
F^{ph,ext}_{\alpha \mu, \beta\nu}(\Omega,\epsilon,\epsilon')
\nonumber && = \sum_{i} \biggl[ \frac{(\gamma^{\mu;i}_{\beta})^*
\gamma^{\nu;i}_{\alpha}} { \epsilon - \epsilon'+
(\Omega_i-i\delta)}\nonumber \\
&&-\frac{(\gamma^{\alpha;i}_{\nu})^* \gamma^{\beta;i}_{\mu}} {
\epsilon - \epsilon'- (\Omega_i-i\delta)} \biggr]\,.
\end{eqnarray}
This model has been applied, for example, to M1 resonances in
closed shell nuclei \cite{KAM89}. In general, the approximation
has problems with the so-called second-order poles, which give
rise to a distortion of the strength function near these poles.
For this reason the model has been extended in several steps: (I)
summation of the $g^2$ terms \cite{KT86} and (II) partial
summation of diagrams \cite{VT89}, which is called
$\emph{chronological decoupling of diagrams}$. In the latter case
all 1p1h$\otimes$phonon contributions are consistently included
and all more complex configurations are excluded---as long as one
neglects ground state correlations. The actual formulas, however,
include the ground state correlations completely.

In the present analytical approach, as in \cite{VT89}, two types
of ground state correlations are included: the conventional RPA
ground state correlations, which affect the location and the
magnitude of the residua of the ph propagators only; and the new
type of ground state correlations, caused by the phonons. These
correlations are qualitatively different from the conventional RPA
correlations because they create new poles in the propagator,
which then cause transitions between the 1p1h$\otimes$phonon in
the ground state and the excited states. They give rise to a
qualitative change of the strength distribution and a change in
the sum rules for the moments of the strength function. In the
present numerical application we only included the conventional
RPA correlations. Calculations which include the complete ground
state correlations are in progress.

As in the TFFS, the energy dependence of the generalized
propagator, $A^{ext}$, is much stronger than that of the
generalized ph interaction, as the next-to-leading order energy
dependence is removed from the interaction and explicitly taken
into account in the generalized propagator. Therefore one
considers explicitly only the energy dependence of the propagator
and neglects the energy dependence of the ph interaction. The
interaction is parameterized as before, although the corresponding
parameters may differ from the previous ones. The final equation
for the change of the quasi-particle density in an external field
has a form identical to that in the TFFS (eq.(\ref{eq:45})):
\begin{eqnarray}\label{eq:51}
&&{\widetilde{\rho}^{ext}(\bm{r},\Omega) =-\int
d^{3}\bm{r}'A^{ext}(\bm{r},\bm{r}',\Omega)}
Q^{eff}(\bm{r}',\Omega)\\ \nonumber &&{-\int
d^3\bm{r}'d^3\bm{r}''A^{ext}(\bm{r},\bm{r}',\Omega)}
F^{ph,}(\bm{r}',\bm{r}'')
\widetilde{\rho}^{ext}(\bm{r}'',\Omega)\,.
\end{eqnarray}
The analytic form of the generalized propagator can be found in
Ref. \cite{rev}.

In the following we investigate, within the continuum random phase
approximation (CRPA) and the ETFFS, electric dipole states in
$^{208}$Pb. The strength distributions for isovector as well as
isoscalar transitions are shown. The single-particle continuum has
been included at the RPA level where it is taken into account
correctly within  our Green function technique in the coordinate
representation. For details see \cite{rev}. In our approach we
consider therefore the three mechanisms that create the width of
giant resonance, namely (I) the Landau damping ((Q)RPA
configurations), (II) the escape width (the single-particle
continuum) and (III) the spreading width (phonon coupling, or
complex configurations). As in all our previous calculations
within the ETFFS, we include the most collective low-lying
phonons. These phonons have been microscopically calculated within
the RPA. In the present and in  our previous calculations we used
the procedure developed in \cite{kl98} to obtain the energy of the
spurious E1 state to be exactly equal to zero without having to
use the procedure of fitting force parameters.

\subsection{Single-particle basis}

\begin{table}[htb]
\centering
\begin{tabular}{|r|l|}
\hline
L$^{\pi}$ & energies [MeV] \\
\hline\hline
$2^+$ & 4.09\ 6.44 \\
\hline
$3^-$ & 2.61 \\
\hline
$4^+$ & 4.32\ 5.44\ 6.00 \\
\hline
$5^-$ & 3.20\ 3.66\ 5.29\ 6.14\ 7.22 \\
\hline
$6^+$ & 4.42\ 5.21 \\
\hline
$7^-$ & 4.04\ 4.69\ 5.03\ 5.66\ 6.33\ 7.17 \\
\hline
$8^+$ & 4.61\ 4.99\ 5.22\ 6.23\ 6.43 \\
\hline
\end{tabular}
\caption{Phonons used in the ETFFS calculation}
\label{tab:1}
\end{table}
We have seen that in the conventional TFFS Landau's quasi-particle
are the single-particle states of the neighboring odd mass nuclei.
Therefore one has to use as input data in the corresponding
equation for the excited states of the even nuclei (renormalized
RPA) the experimental single-particle energies. These
single-particle energies include obviously the effect of the
coupling to more complicated configurations e.g. phonons. In the
actual calculation within the \emph{extended theory} we consider
about twenty low-lying phonons explicitly and for that reason one
has first to determine \emph{bare} single-particle energies that
do not include the coupling to those phonons. If we couple the
corresponding phonons to the $\emph{bare}$ single-particle states
we reproduce the original quasi-particle energies. In order to
obtain the new \emph{refined} single-particle basis
$\widetilde{\epsilon}_\nu$ one has to solve eq.(\ref{eq:52}), which
can be obtain from eq.{(\ref{eq:48}):
\begin{equation}\label{eq:52}
\widetilde{\epsilon}_{\nu_1} =\epsilon_{\nu_1} -\sum_{\nu_2 ;i}
 \frac{\left| \gamma^{\nu_2 ;i}_{\nu_1}\right|^2}
{\epsilon_{\nu_1} -\Omega_i  -\widetilde{\epsilon}_{\nu_2}}\,.
\end{equation}

Equation ( \ref{eq:52}) has been first derived by Ring and Werner \cite {RW73}.
It has been applied to the calculation of the fragmentation of the single particle strength in the neighboring
odd mass nuclei of $^{208}Pb$ \cite{RW73} and the high-spin states in $^{208}Pb$ Ref. \cite {Krew80}.

\begin{figure}[htb]
\begin{center}
\includegraphics[width=9cm,angle=270]{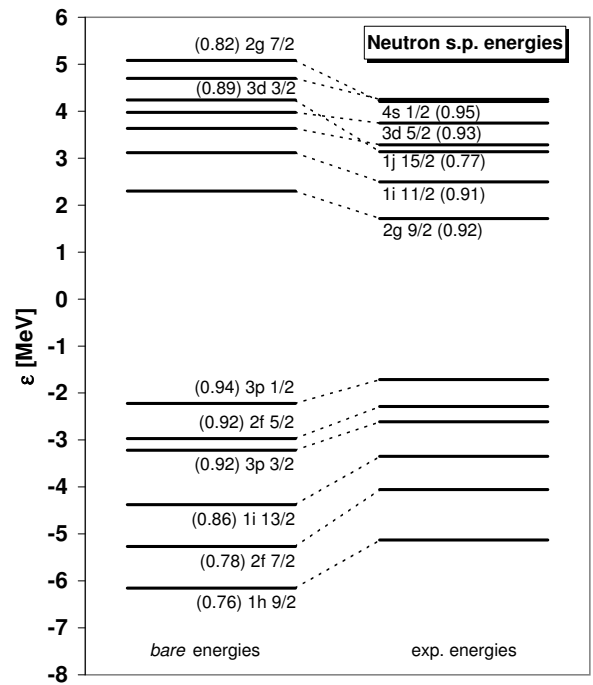}
\end{center}
\caption{\label{fig:3}Neutron single-particle energies in $^{208}$Pb. Here we compare the experimental values with the \emph{bare} single-particle energies which were obtained from eq.(\ref{eq:52}). The values in parentheses are the single particle strengths.}
\end{figure}

\begin{figure}[htb]
\begin{center}
\includegraphics[width=9cm,angle=270]{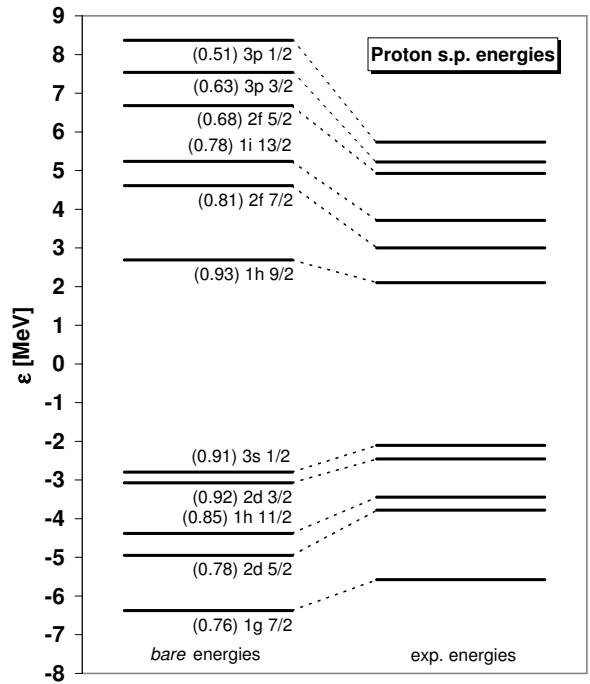}
\end{center}
\caption{\label{fig:4} Same as in Fig. \ref{fig:3} for the proton single-particle energies in
$^{208}$Pb}.
\end{figure}

The phonons that are explicitly considered in the present
calculation are given in Table \ref{tab:1}.
In Figs. \ref{fig:3} and \ref{fig:4} the \emph{bare}
single-particle spectrum for protons and neutrons are shown and
compared with the experimental spectrum. The energy shift due to
the coupling to the phonons is nearly twice as large for the
protons as for the neutrons. In addition, one obtains a
fragmentation of the single-particle strength. The corresponding values are also
given in  Figs. \ref{fig:3} and \ref{fig:4}. Here we point out that, in addition to the phonon
coupling, the short range correlations in the \emph{G-matrix} also
reduce in a uniform way the single-particle strength. This part of
the reduction is taken care of by the renormalization procedure in
section II. The \emph{bare energies} may be identified with the
mean energies that one obtains within a self-consistent approach.
From the present calculation one observes that the (in general)
too large spreading of the self-consistent single-particle spectra
compared with the experimental ones  would be reduced on the
average for the neutrons by about 1 MeV and for the protons by 1.5
MeV with relatively large fluctuations. If one solves equation
(\ref{eq:52}) with the corresponding mean field solutions one
obtains the \emph{dressed} single-particle energies that should
agree (from our point of view) with the experimental spectrum. The
number of phonons can be restricted to the most collective ones
because they are the most important ones for the width of the
giant resonances. They also influence the different
single-particle states in an individual manner. If we increase the
number of phonons one only obtains an overall shift that is
equivalent to a change in the effective mass. We conclude that for
a given set of phonons one has to chose an effective interaction
with the appropriate effective mass and spin orbit interaction. We
return to this question in the final discussion.

\subsection{The residual particle-hole interaction}

As in our previous calculations we used the effective
particle-hole Landau-Migdal interaction
\begin{eqnarray}\label{eq:53}
&&F(\bm{r},\bm{r'}) = C_{0}\left[f(r) + f'( r)\bm{\tau}_1
\cdot\bm{\tau_2}  \right. \nonumber \\
&& +\left.(g + g' \bm{\tau_1}\cdot\bm{\tau_2})\bm{\sigma_1}\cdot\bm{\sigma_2} \right]
\;\delta(\bm{r}-\bm{r'})
\end{eqnarray}
with the conventional interpolation formula, for example, for the
parameter $f$,
\begin{equation}\label{eq:54}
f(r) = f_{ex} + (f_{in} - f_{ex})\rho_{0}(r)/\rho_{0}(0)
\end{equation}
and similarly for the other $r$-dependent parameter $f'(r)$. Here
$\rho_{0}(r)$ is the density distribution of the ground state of
the nucleus under consideration and $f_{in}$ and $f_{ex}$ are the
force parameters inside and outside of the nucleus. The standard
values of the parameters, which have been used for all the nuclei
under consideration \cite{rev}, are:
\begin{eqnarray}\label{eq:55}
& f_{in}  =  - 0.002,\; f_{ex} =-1.35,\;
f_{ex}^{\prime}  =  2.30, \; f_{in}^{\prime}  =  0.76, \nonumber \\
& g_{in} = g_{ex} =  0.05,\;g^{\prime}_{in}  = g^{\prime}_{ex}= 0.96.
\end{eqnarray}
The parameters are given in units of $C_{0}  =  300\;{\rm
MeV fm^{3}}$. For the nuclear density $\rho_{0}(r)$ in the
interpolation formula we chose the theoretical ground state
density distribution of the corresponding nucleus,
\begin{equation}\label{eq:56}
\rho_{0}(r) = \sum_{\epsilon_{i} \leq \epsilon_{F}} 
\frac{1}{4\pi}(2j_{i} + 1)   R^{2}_{i} (r)\,.
\end{equation}
Here  $R_{i}(r)$ are the single-particle radial wave functions of
the Woods-Saxon single-particle model. This is more consistent
than the previously used phenomenological Fermi distribution. For
that reason we readjusted in the present calculation $f_{ex}$ to reproduce the breathing mode and obtained the value of
$f_{ex}$ = -1.35. For other details of the
calculations, see Ref.\cite{rev,GS06}.

\section{Low-lying electric dipole strength}

\subsection{The isovector case}

The giant electric dipole resonance (GDR) is one of the most collective states in nuclei. 
It exhausts a major part of the energy-weighted sum rule; the excitation energy is a smooth
function of the mass number; and the form of the GDR changes only little from nucleus to nucleus. 
Phenomenological collective models describe well the A-dependence of the energy and the strength 
with only a few parameters. 
In microscopic models these collective states are coherent superpositions of many particle-hole 
states.
\begin{figure}[htb]
\begin{center}
\includegraphics[width=9cm]{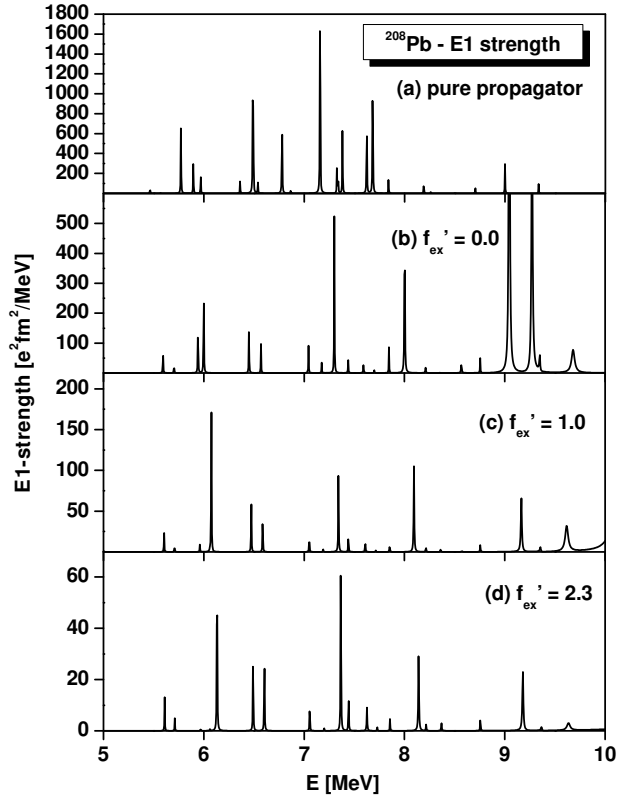}
\end{center}
\caption{\label{fig:5} Electromagnetic E1 strength distribution calculated
within the conventional 1p1h RPA. In (a) we show the uncorrelated
ph-matrix elements. In (b) we used our universal isoscalar
interaction and the spurious state has been removed. For the
results shown in (c) a reduced isovector force has been used,
while (d) shows the results with the full interaction.}
\end{figure}
The theoretical results of such models for the mean energy and the
total strength agree, in general, very well with the data. The
strength distribution, however, is reproduced by only a few very
involved models, such as the ETFFS considered here.

Here we first investigate the question whether the pygmy dipole
resonances (PDR) are like the GDR collective states in the
microscopic definition. That is, are they coherent superpositions
of many particle-hole states? In order to answer this question we
performed a series of calculations within the continuum RPA in
which we varied the particle-hole interaction.

In Fig.\ref{fig:5}a the dipole distribution for the
electromagnetic $E1$-operator is plotted, where the particle-hole
interaction is zero; i.e. the figure shows pure proton and neutron
ph-matrix elements. The largest matrix elements are the ones
between the ph states with the largest angular momenta that differ
by one unit---the \emph{stretched configurations}. The state at
$E\approx $6.5 MeV is the neutron $(1j_{15/2})(1i_{13/2})^{-1}$
configuration and the states at $E\approx$ 7.15 MeV and $E\approx
$ 7.70 MeV are the proton configurations
$(1i_{13/2})(1h_{11/2})^{-1}$ and $(1h_{9/2})(1g_{7/2})^{-1}$. 
The three largest transitions exhaust about 40\% of the total strength. 
Here, we have to bear in mind that the transitions shown also include spurious components that 
give rise to the spurious isoscalar state at zero energy. 
This state corresponds to the translation of the whole nucleus. 
For that reason one has to remove the spurious components in each of the particle-hole
configurations. 
The method developed in Ref. \cite{kl98} is used that brings the spurious state exactly to zero 
energy and removes completely the spurious strength in all excited states.

In Fig.\ref{fig:5}b-d the electromagnetic dipole strength, from
which the spurious strength has been removed, is shown.

\begin{figure}[htb]
\begin{center}
\includegraphics[width=9cm]{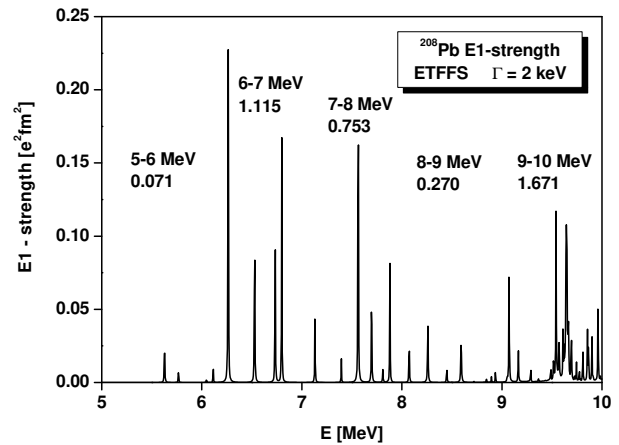}
\end{center}
\caption{\label{fig:6} Integrated electromagnetic E1 strength calculated
within the ETFFS. These results should be compared with the
distribution shown in Fig.\ref{fig:5}d.  Note that the strength in
the present figure is integrated over the smearing width, given in $\rm e^{2}fm^2$.}
\end{figure}

To demonstrate the effect of the repulsive isovector interaction
we have chosen three different values for the isovector force
parameter $f'_{ex}$. The isoscalar ph interaction is the same for all
three cases. It reproduces the breathing mode, which is later shown in
Fig.\ref{fig:20}. If we compare the completely uncorrelated
strength distribution in Fig.\ref{fig:5}a with the results in
Fig.\ref{fig:5}b, where the spurious components are removed but
the isovector force parameter $f'_{ex} =0$, one observes an
appreciable shift of the B(E1) strength to higher energies. We
notice thereby that if one removes the spurious strength one has
to introduce correlations that give rise to this redistribution of
the strength. With setting the isovector interaction to zero, most
of the B(E1) strength (87\%) lies below 10 MeV. If we increase the
isovector interaction parameter to $f'_{ex} = 1.0$
(Fig.\ref{fig:5}c), which is about half of the force parameter
that reproduces the experimental data of the GDR, the dipole
strength below 10 MeV is reduced to $11\%$.

\begin{figure}[htb]
\begin{center}
\includegraphics[width=9cm]{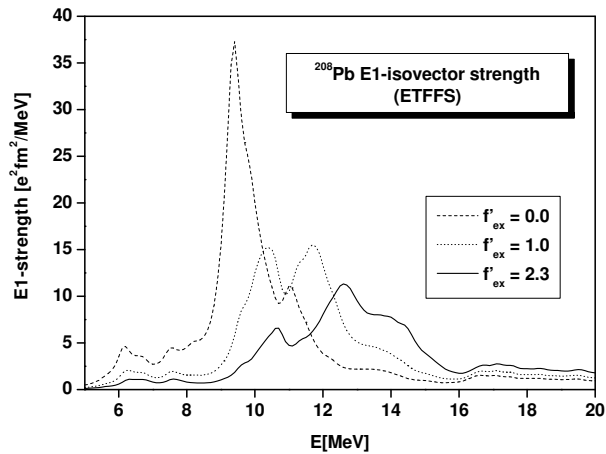}
\end{center}
\caption{\label{fig:7} Electromagnetic E1 strength from 5 to 20
MeV calculated in our \emph{extended theory}. The thick line
indicates the final result whereas the other lines are the results
of calculations where the isovector force parameter has been
varied as indicated in the figure.}
\end{figure}

Finally, in Fig.\ref{fig:5}d, where we use our conventional force
parameter (which reproduces nearly quantitatively the experimental
data of the GDR), only $5.2\%$ of the strength remains below 10
MeV. We observe that the major part of the E1 strength is shifted
into the GDR region, where it creates a very collective
resonance---in agreement with the data.

Within our model, in the electromagnetic E1 strength up to 9 MeV,
which is the region of the \emph{pygmy resonance}, we do not find
one single state in which most of the low-lying strength is
concentrated, but we find already in the 1p1h RPA several states
whose energies are essentially unchanged compared with the $f'_{ex}
=0$ case. The E1 strength, however, is reduced by a factor of 10
due to the strongly repulsive isovector force, which shifts the
strength into the GDR region and produces a collective resonance.
Such a behavior is expected from the schematic model of Brown and
Bolsterli \cite{GEB}.

In Fig.\ref{fig:6} we present the results of the ETFFS calculation
where, in addition to the single-particle continuum, the effect of
the phonons is also included. The phonons that we consider are
given in Table \ref{tab:1}. The result in Fig.\ref{fig:6} should
be compared with the results in Fig.\ref{fig:5}d. One observes a
further fragmentation of the strength and a small shift of the two
strongest states to higher energies. Whereas the calculated
strength distribution between 7 and 8 MeV is in good agreement
with the data, we obtain too little strength below 6 MeV and too
much between 6-7 MeV as compared with the present data. A shift of
the two neutron states $1j_{15/2}$ and $(1i_{13/2})^{-1}$  improves
the agreement between theory and experiment.

In Fig.\ref{fig:7} the electromagnetic E1 strength distribution
calculated within the ETFFS is shown up to 20 MeV with a
smearing parameter $\Delta$ = 250 keV and three different
$f'_{ex}$ parameters. It is interesting to see how the strength is
shifted to higher energies with increasing interaction strength.
For our conventional value $f'_{ex}$ = 2.30 the theoretical
distribution in the GDR region agrees quantitatively with the
experimental data. In connection with the previous discussion, the
energy range between 6-8 MeV is of special interest. With no
isovector interaction, but the spurious components removed, one
obtains some concentration of strength around 6.2 MeV and 7.5 MeV.
The latter is considered as the PDR region. With increasing
isovector force this strength is reduced. For the full force the
reduction is a factor of 5 compared to $f'_{ex} = 0$. This behavior
is just the opposite of a collective structure, where with
increasing force the collectivity is enhanced.

\begin{figure}[htb]
\begin{center}
\includegraphics[width=9cm]{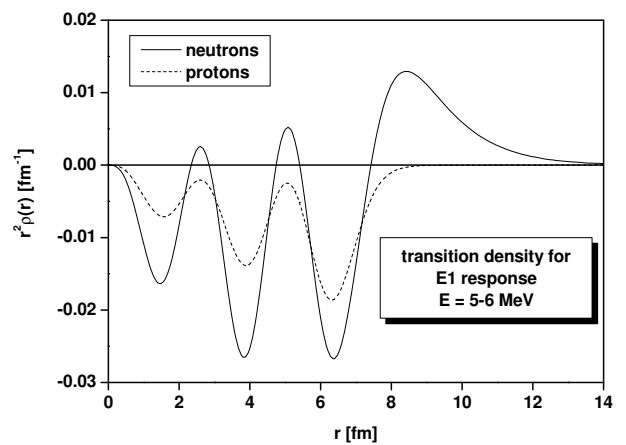}
\end{center}
\caption{\label{fig:8} Summed transition densities for the
electric dipole response shown in Fig.\ref{fig:6} from 5 to 6 MeV}
\end{figure}

\begin{figure}[htb]
\begin{center}
\includegraphics[width=9cm]{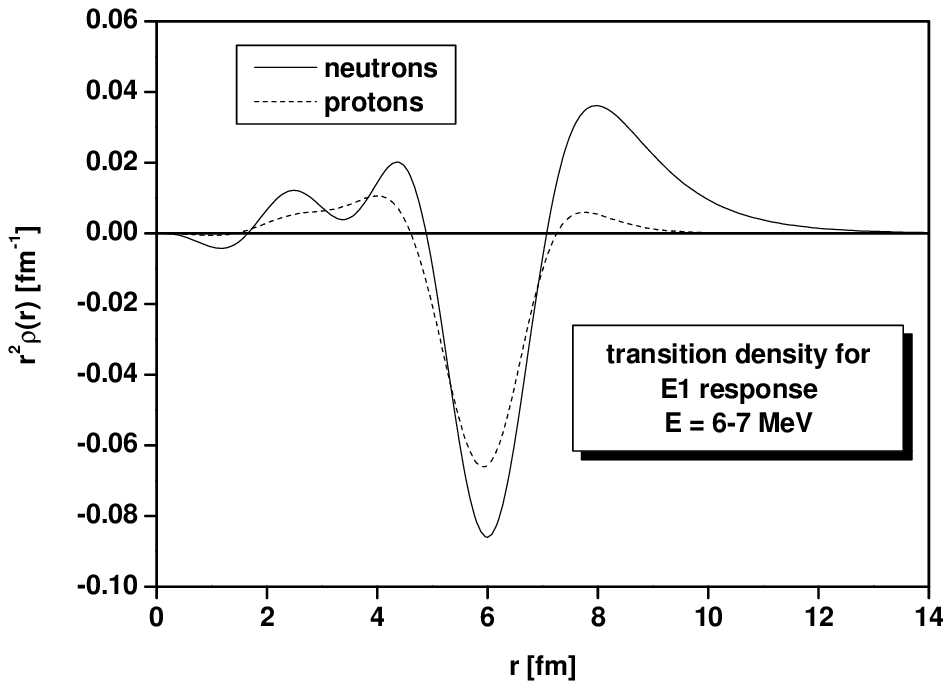}
\end{center}
\caption{\label{fig:9} Summed transition densities for the
electric dipole response shown in Fig.\ref{fig:6} from 6 to 7 MeV}
\end{figure}

\begin{figure}[htb]
\begin{center}
\includegraphics[width=9cm]{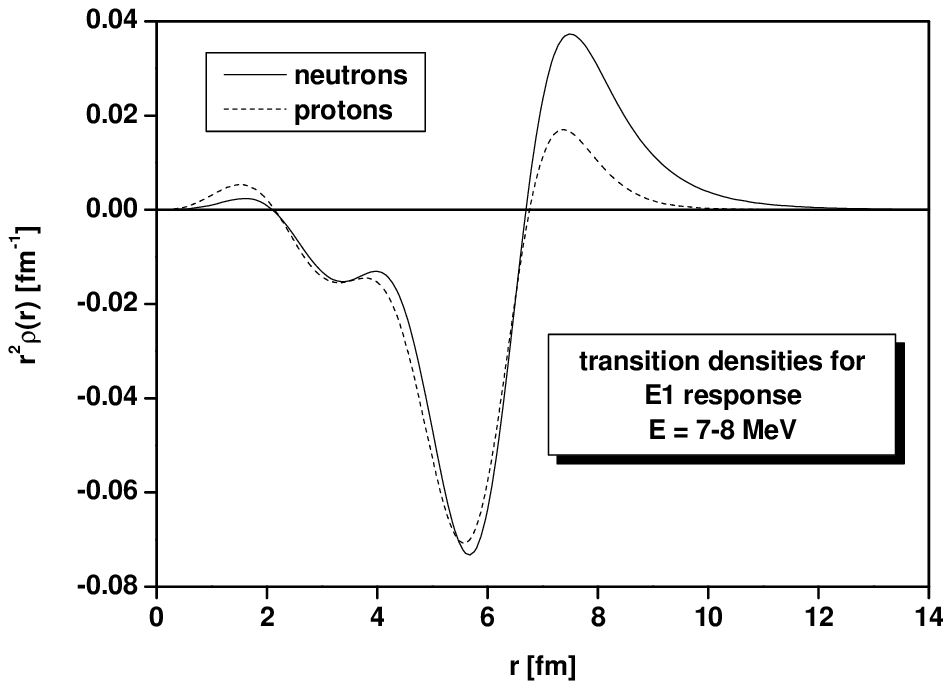}
\end{center}
\caption{\label{fig:10} Summed transition densities for the
electric dipole response shown in Fig.\ref{fig:6} from 7 to 8 MeV}
\end{figure}

In Figs.\ref{fig:8}-\ref{fig:10} the transition densities between
5-6 MeV, 6-7 MeV and 7-8 MeV of the electric dipole response are
plotted for transitions shown in Fig.\ref{fig:6}. In all three
cases the protons and neutrons are in phase, which indicates
strong isoscalar components. This at-first-glance surprising
behavior has also been found by other authors. It can be
understood immediately in our microscopic model. As discussed in
the next section, the isoscalar force is attractive and gives rise
to a collective structure around 22 MeV. Due to this attractive
force, some of the high-lying strength couples also to the
low-lying states, whereas the strong repulsive isovector force
simultaneously shifts the isovector strength from these
configurations into the GDR region. We demonstrate this behavior
in {Figs.\ref{fig:12}-\ref{fig:14} where the transition densities
are plotted for the same energy ranges but with the force
parameters $f'_{ex}$ set to zero. As the isovector force is zero, the
isovector strength remains in the low energy region, which can be
clearly seen in the transition densities.

\subsection{The isoscalar case}

As mentioned earlier, for the electric isoscalar dipole states we
have a special situation because the lowest isoscalar resonance is
the spurious state corresponding to the translation of the whole
nucleus. In self-consistent calculations this state appears (at
least in principle) at zero energy and carries all the spurious
strength.

\begin{figure}[htb]
\begin{center}
\includegraphics[width=9cm]{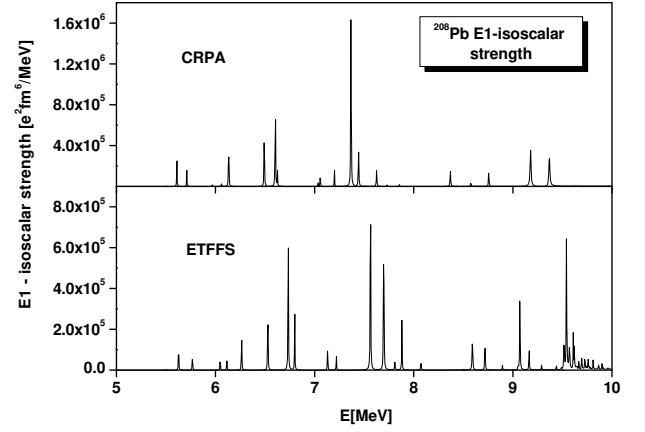}
\end{center}
\caption{\label{fig:11} E1 isoscalar strength distribution in
$^{208}$Pb from 5-10 MeV for the transition operator 
$(r^3-5/3\langle r^2 \rangle r) Y_{1,\mu}$. 
Here we compare the results derived from the CRPA with the results obtained with the
\emph{extended theory}.}
\end{figure}

\begin{figure}[htb]
\begin{center}
\includegraphics[width=9cm]{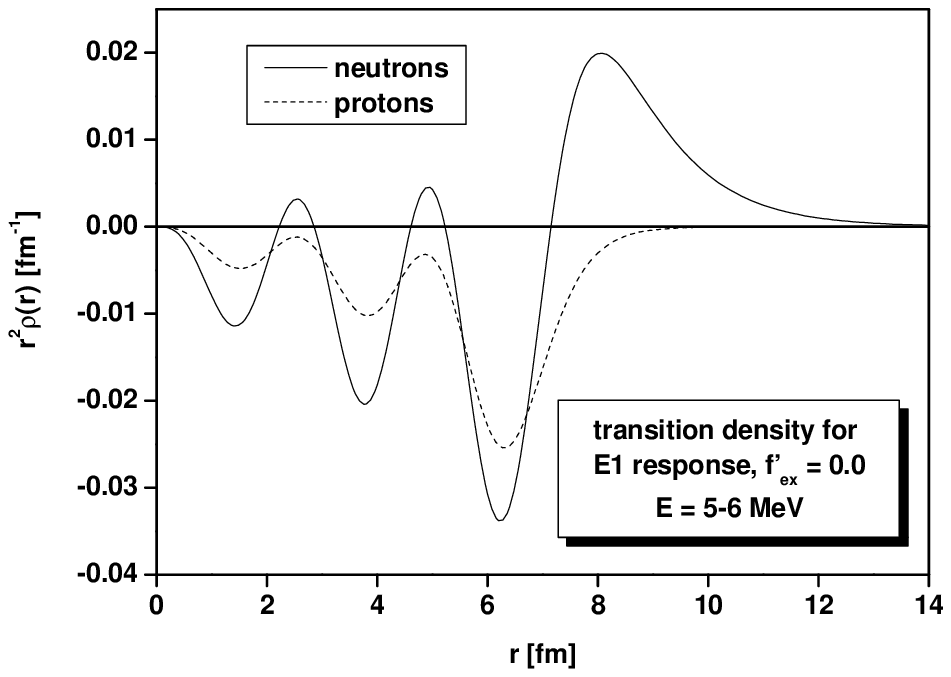}
\end{center}
\caption{\label{fig:12} Summed transition densities from 5 to 6
MeV with the isovector force parameter $f'_{ex} =0$ }
\end{figure}

\begin{figure}[htb]
\begin{center}
\includegraphics[width=9cm]{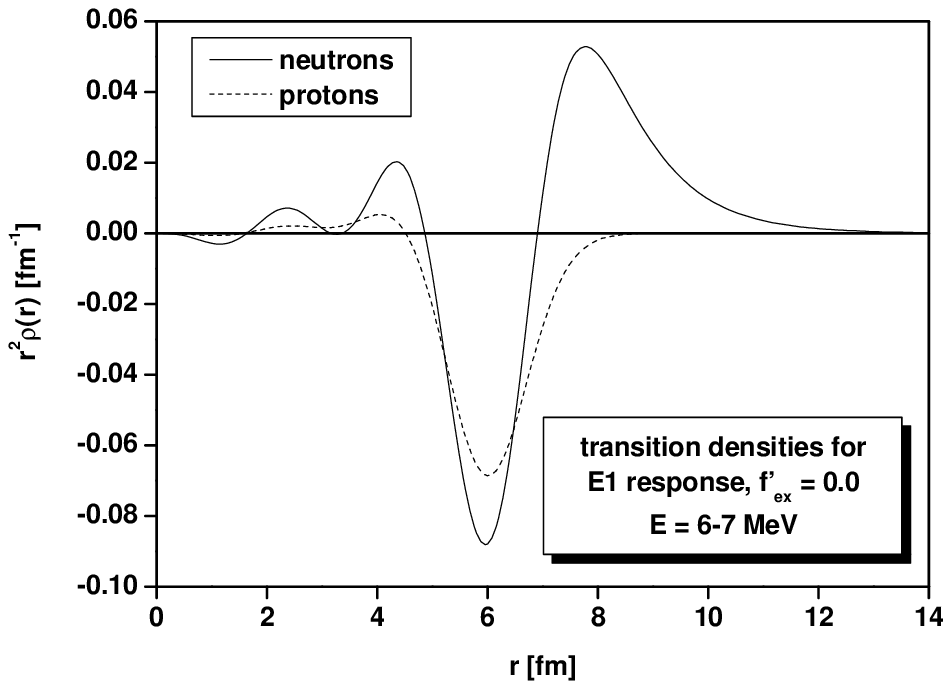}
\end{center}
\caption{\label{fig:13} Summed transition densities from 6 to 7
MeV with the isovector force parameter $f'_{ex} =0$}
\end{figure}

\begin{figure}[htb]
\begin{center}
\includegraphics[width=9cm]{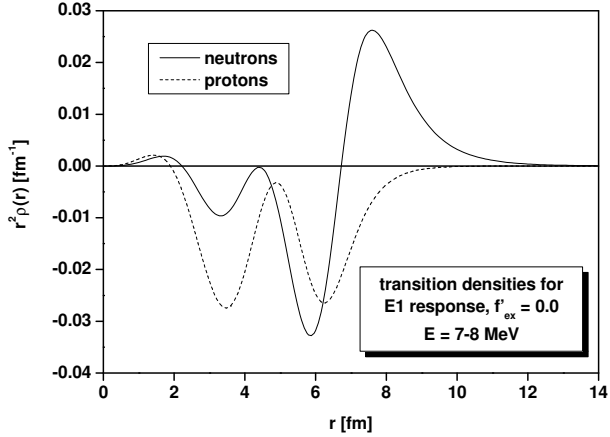}
\end{center}
\caption{\label{fig:14} Summed transition densities from 7 to 8 MeV with the isovector force
parameter $f'_{ex} =0$}
\end{figure}

In the Landau-Migdal approaches the spurious state has to be removed explicitly. 
Here we use the procedure developed in Ref.\cite{kl98} to obtain the energy of the spurious
E1 state exactly at zero energy, with no spurious components in the excited state. 
As the isoscalar electric dipole strength
distribution depends on the isoscalar parameter $f$, we test the
force by calculating the isoscalar monopole resonance---the
breathing mode. With our universal value of $f_{ex} = -1.35$ we
reproduce the mean energy as well as the width of the resonance
(see the next section). 
In Fig.\ref{fig:11} the results of our calculation for the isoscalar dipole operator 
$(r^3-5/3\langle r^2 \rangle r) Y_{1,\mu}$ up to 10 MeV are presented. 
There we compare the CRPA with the ETFFS results. 
In the CRPA we obtain one´strong state near 7.4 MeV and several somewhat weaker states at
lower energies. 
Due to the phonon coupling the strength of the strongest state is fragmented and shifted to 
slightly higher energies. 
Those isoscalar states have been investigated experimentally with the 
$(\alpha,\alpha',\gamma_0)$ reaction. 
So far the experimental data from $^{208}$Pb are not given in the
form of cross sections but only in counts \cite{hvw01}. The data
show a very strong signal at about 5.6 MeV and a somewhat weaker
signal around 6.7 MeV. As the isoscalar strength distribution in
Fig.\ref{fig:11} is calculated with the operator $(r^3-5/3\langle r^2 \rangle r) Y_{1,\mu}$, a
comparison with these data is not directly possible because the
$(\alpha,\alpha')$ reaction is sensitive to the tails of the
isoscalar distributions and the $\gamma_0 $ is proportional the
isovector admixture. Unfortunately, analyzing programs in which
microscopic transition densities can be used do not yet exist
\cite{Zil06}. The transition densities shown previously are the
same for the isoscalar and isovector case. Only the transition
strength is different because the corresponding operators weight
the transition densities differently.

\section{High-lying electric dipole strength}
\begin{figure}[htb]
\begin{center}
\includegraphics[width=9cm]{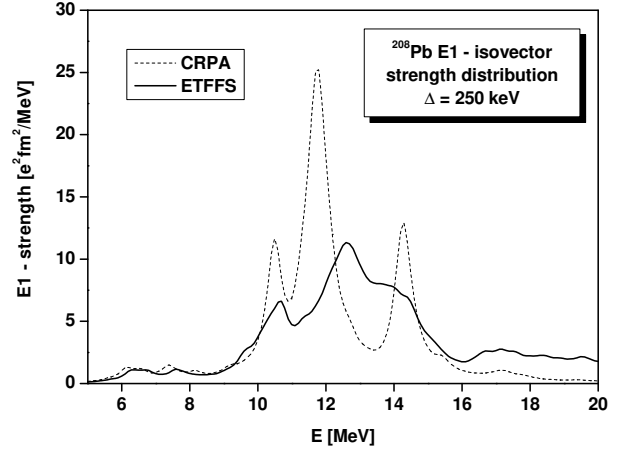}
\end{center}
\caption{\label{fig:15} E1 isovector strength from 5 to 20 MeV. We
compare here the results of the continuum RPA with the
\emph{extended theory} the width and the mean energy of the latter
agree quantitatively with the experimental results.}
\end{figure}

\begin{figure}[htb]
\begin{center}
\includegraphics[width=5cm,angle=90]{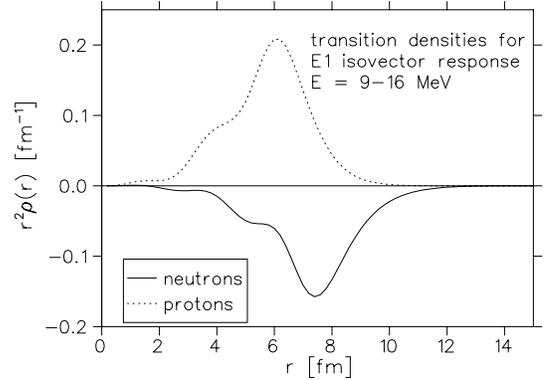}
\end{center}
\caption{\label{fig:16} Transition density of the isovector giant
dipole resonance}
\end{figure}

\begin{figure}[htb]
\begin{center}
\includegraphics[width=9cm]{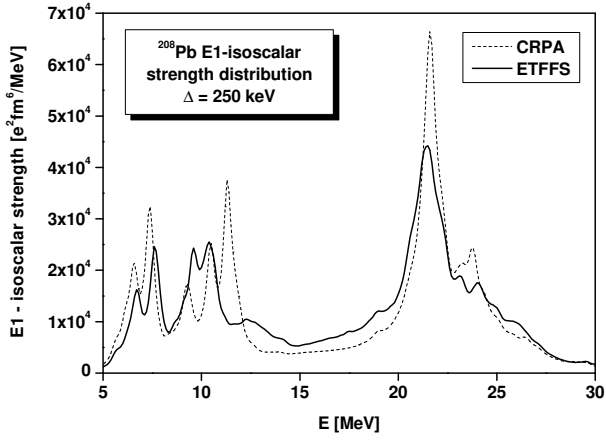}
\end{center}
\caption{\label{fig:17} Isoscalar E1 strength from 5 to 30 MeV calculated with
a smearing parameter $\Delta$ = 250 keV}
\end{figure}

\begin{figure}[htb]
\begin{center}
\includegraphics[width=5cm,angle=90]{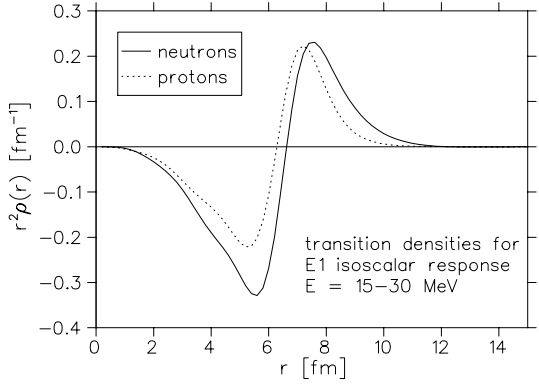}
\end{center}
\caption{\label{fig:18} Transition density of the isoscalar giant
dipole resonance}
\end{figure}

\begin{figure}[htb]
\begin{center}
\includegraphics[width=5cm,angle=90]{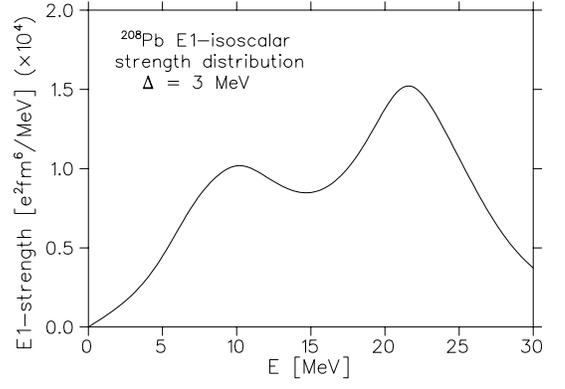}
\end{center}
\caption{\label{fig:19} Isoscalar E1 strength  calculated with a smearing parameter
$\Delta $= 3 MeV. Note: The larger smearing parameter is justified only for the high-lying 
($\geq $15 MeV) part of the resonance.}
\end{figure}

\begin{figure}[htb]
\begin{center}
\includegraphics[width=9cm]{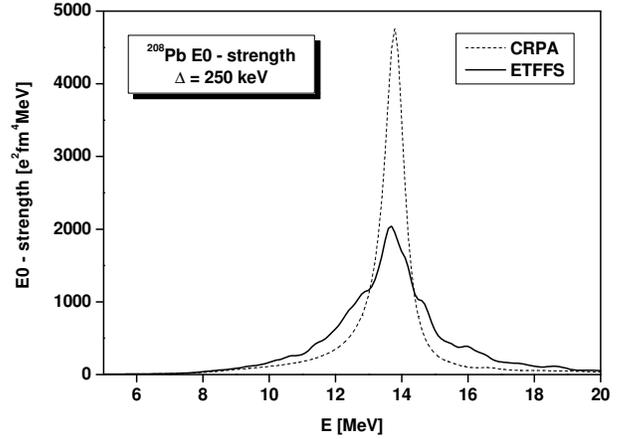}
\end{center}
\caption{\label{fig:20} Isoscalar E0 strength from 5 to 20 MeV
calculated in the continuum RPA and the \emph{extended theory}.
The mean energy and width of the latter agrees with the data.}
\end{figure}

\begin{figure}[htb]
\begin{center}
\includegraphics[width=5cm,angle=90]{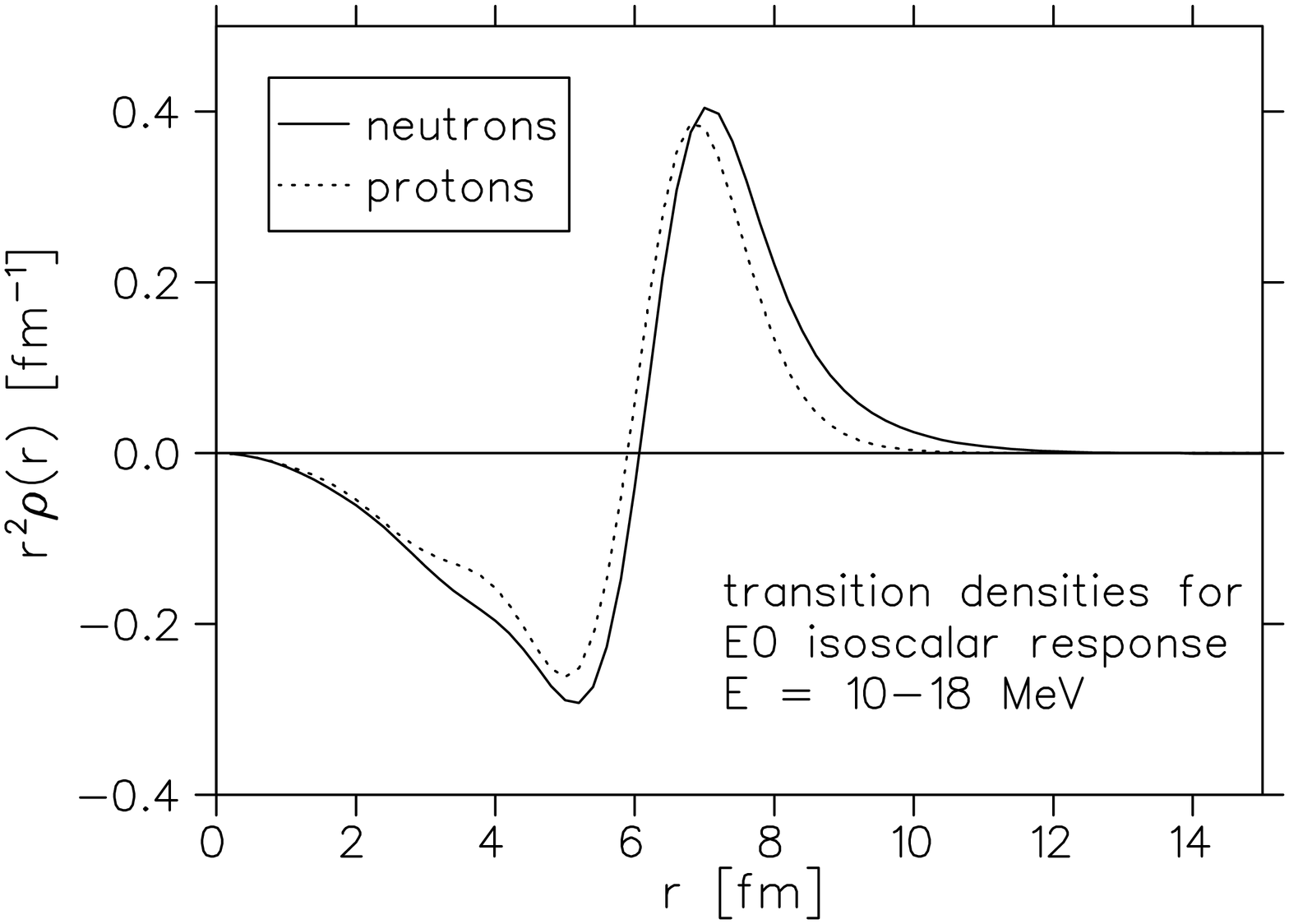}
\end{center}
\caption{\label{fig:21} Transition density of the breathing mode}
\end{figure}

The GDR was calculated with our universal parameters given in eq.({\ref{eq:55}})
and the result is shown in Fig.\ref{fig:15}. The mean energy
and the width of the resonance agree with the data within the
error bars. The escape and spreading widths are included in our
microscopic model. In addition, we consider a \emph{smearing}
parameter $\Delta$ = 250 keV , which corrects for more complex
configurations that have not be considered in our approach. Here
we point out again that the low-lying and high-lying E1 strength
is calculated within the same model with exactly the same
single-particle energies and force parameters. This differs from
some \emph{self-consistent} approaches in which the
single-particle spectrum is modified in order to fit the low-lying
spectrum e.g. \cite{HL02}.

In Fig.\ref{fig:16} the corresponding transition density is plotted. It is
surface-peaked and protons and neutrons are out of phase. As we
have 50\% more neutrons than protons, the neutron density has not
only a stronger and longer tail than the proton density, but it is
also peaked further out. For that reason we also have some
admixture of isoscalar E1 strength---even in the isovector giant
dipole region. The form of the transition density agrees
with surface-peaked phenomenological models.

In Fig.\ref{fig:17} the theoretical isoscalar E1 response from 5-30 MeV is
plotted. One realizes that, in addition to the low-lying strength,
we obtain a strong collective structure above 20 MeV, which
represents the isoscalar electric dipole resonance. The mean energy of the resonance is
$\widetilde{E}_{th}$ = 22.1 MeV and the width $ \Gamma_{th}$ = 3.8 MeV if we chose a
smearing parameter $\Delta$ = 250 keV. 
The deep-lying holes contribute a major part to the high-lying E1-strength. 
These holes are several MeV broad. 
These widths are not included in the calculation shown in Fig.\ref{fig:17} where we used a smearing 
parameter of $\Delta$ = 250 keV. 
In order to correct for these widths we performed a calculation with $\Delta$ = 3 MeV, 
the corresponding result is shown in Fig.\ref{fig:19}. 
The theoretical results can be compared with a recent $(\alpha, \alpha')$ experiment 
\cite {Uchida03}, which obtained $ \widetilde{E}_{exp}$ = 22.5-23.0 MeV and a width 
$\Gamma_{exp} \approx$ 10 MeV.
The corresponding transition density is given in Fig.\ref{fig:18}. 
It has the form of a compression mode and for that reason we have 
shown---finally---in Figs.\ref{fig:20} and \ref{fig:21} the isoscalar monopole strength 
and the corresponding transition density. 
These calculations are performed for purpose of
consistency because the isoscalar monopole and isoscalar dipole
depend both on the isoscalar parameters $ f $ and have, in
addition, a very similar structure, as can be seen in the
transition densities. In the monopole case the radial integral
over the transition density has to be zero; in the electric dipole
case the transition density multiplied by the radial coordinate
$r$ must vanish.

\section{Discussion}

The center of our investigations has been the origin and the structure of the pygmy resonances. 
We have chosen in $^{208}$Pb a nucleus with an excess of 44 neutrons, which means that we have
$50\%$ more neutrons than protons. 
The structure of this doubly-magic nucleus, as well as of the neighboring odd mass nuclei, is well 
known and all microscopic models and microscopic 
theories---self-consistent ones and Landau-Migdal ---work best for this case. 

Our calculation shows, first of all, that we obtain isovector as well as isoscalar electric dipole 
states below 10 MeV. 
This is obvious, as the ph-interaction has isoscalar as well as isovector components. 
Here we point out that all the spurious components are removed; i.e. the expectation value of the
translation operator $ rY_{1\mu}$ in all the excited states is negligibly small. 
The isoscalar strength is due to the $3\hbar \omega$ and higher ph components. 
In the present Landau-Migdal approach one is able to remove the spurious strength without
performing a full RPA calculation, as must be done in self-consistent approaches. 
Here, the force parameters can be varied independently. 
This allows one to investigate the influence of the isovector interaction on the low-lying 
isovector dipole strength in detail. 

We have shown in Figs.5 and 7 that if one starts with the experimental single-particle energies, 
then with increasing isovector force the isovector dipole strength is shifted to higher energies 
and one is finally left with a small fraction of the sum rule in a few states that are only 
slightly shifted with respect to the unperturbed ph energies. 
There is no indication in the present model of a low-lying collective state that is built up by 
the isovector interaction. 
The theoretical electromagnetic strength below 8 MeV is 1.94 $[ e^2fm^2]$ compared with the 
experimental strength of 1.32 $[ e^2fm^2]$. 
The distribution of the theoretical strength agrees in the range between 7-8 MeV with the current 
data. 
The calculated strength between 6-7 MeV, however, is too large and the strength below 6 MeV too 
small compared with the experimental values. 
On the other hand, the width and the mean energy of the GDR is reproduced quantitatively in our 
model.

It is obvious that if one changes the single-particle spectrum one is always able to improve the 
agreement with the data. 
In the present calculation we only consider the lowest order in the expansion of the Migdal 
interaction, as is done in nearly all calculations. 
The next-to-leading order in Migdal's interaction depends on the momenta through 
$ \bm{p}\cdot \bm{p'}$, with the isoscalar force parameter $f_1$ and the isovector force parameter
$f'_1$. 
The sign and magnitude of the two parameters can be estimated from the effective mass of the 
nucleus and the orbital part of the effective magnetic operator \cite{rev77}. 
From this consideration it follows that both parameters are negative, with the $f'_1$ parameter 
slightly smaller in magnitude. 
These parameters would give rise to an attractive velocity dependent isoscalar and isovector ph 
interaction. 
A small effective mass, as is used in relativistic RPA calculations, would increase the magnitude 
of both parameters. 
A strong isovector, velocity dependent force could be the explanation of the collective states 
with large vorticities that are found in relativistic RPA calculations \cite{ring05}. 
In this approach two phonon states will give rise to a fragmentation that should explain the E1 
data between 5 and 9 MeV. 
Unfortunately, the lowest two phonon states are at 6.7 and 6.9 MeV, whereas the lowest 
experimental E1-states are more then one MeV lower. 
We therefore consider the present result, where the E1 strength between 5.5-9 MeV lies essentially 
at the unperturbed ph-energies and the major part of the isovector E1 strength is shifted into the 
GDR, to be the more physical one. 
Moreover, a collective phenomenon has not been found  in self-consistent calculations in which one 
starts with effective forces of the Skyrme type. 
The next-to-leading order in Migdal's interaction ($f_1$,$f^\prime_1$) 
will lead to some redistribution of the low-lying strength which may 
improve the agreement with the data below 7 MeV. 
Such calculations are in progress.

\section{Consequences for self-consistent calculations}

The present calculation is not self-consistent in the sense that one starts with an effective 
Lagrangian (or Hamiltonian), the numerous parameters of which are, in general, adjusted to gross
properties of nuclei and can be used for all nuclei. 
These parameters cannot be determined in a unique way and therefore there exist numerous sets of 
parameters that reproduce the gross properties equally well, but give quite different 
results for excited states. 
If we consider, as a specific example, effective forces of the Skyrme type, we find 
parameterizations that, for example, give rise to effective masses from ${m^*}/{m}$ = .6 to 
${m^*}/{m}$ = 1. 
For nuclear structure calculations the effective mass is of special importance because the spacing 
of the single-particle spectrum is inversely proportional to it; a smaller $m^*/m$ expands the 
single-particle spectrum and a larger one compresses the spectrum. 
The experimental single-particle spectra in the neighboring odd mass nuclei of $^{208}$Pb 
correspond to an effective mass of roughly one, whereas for medium mass nuclei it is smaller then 
one. 
If one describes the spectrum in the odd mass nuclei correctly one simultaneously reproduces the 
non-collective states in $^{208}$Pb. 
The collective states, on the other hand, depend not only on the ph-spectrum but also on the 
residual ph interaction. 
It seems always possible to find, under the numerous sets of Skyrme parameters, one set that 
reproduces the collective states in which one is interested, whereas the non-collective ones 
deviate from the data by several MeV.

In our calculations we investigated simultaneously the E1 spectrum between 5-8 MeV and the 
high-lying spectrum. 
Our calculations show that the E1 spectrum below 8 MeV is very similar to the unperturbed 
experimental ph spectrum, whereas the high-lying collective states depends sensitively on the 
(universal) Migdal parameters. 
This is not an exception; rather, it is the general experience. 
In the low energy spectrum of $^{208}$Pb one finds some collective states with natural parity that 
are collective (e.g. $ 3^-,5^-,2^+ $ etc.) but the majority of the 
states---especially the low-lying unnatural ones---are energetically at the experimental 
ph energies. 
Therefore, if one is looking for a unified description of the structure of even-even nuclei, one 
has also to consider the spectra of the neighboring odd mass nuclei.

\section{Summary}

The \emph{extended theory of finite Fermi systems} has been
applied to the low-lying and high-lying E1 spectrum of $^{208}$Pb
as an example of a neutron-rich nucleus. In the present approach
the low-lying electromagnetic E1 spectrum is non-collective in the
sense that the major part of the strength is not concentrated in
one single state. The strength is rather distributed over several
states which are close to unperturbed ph energies and we find in
the low-lying E1 states strong isoscalar admixtures. The
theoretical results for the GDR as well as the breathing mode
agree quantitatively with the data. We also find in our
calculation a well-localized isoscalar E1 resonance at  22.2 MeV,
with a width of 9.3 MeV. This resonance has been detected in
$(\alpha,\alpha')$ scattering.

From the present investigation one may draw important conclusions
concerning self-consistent calculations. We have seen that the
inclusion of the phonons gives rise to a shift of the
single-particle spectrum of about 1 MeV for neutrons and nearly 2 MeV for protons. 
In the present \emph{extended theory} one has to
determine \emph{bare } single-particle energies that give the
experimental ones if one includes the phonons. 
Self-consistent approaches determine the \emph{bare} spectrum that we need to use
as input in our \emph{extended theory}. 
One therefore has to chose a parametrization of the effective intercation that reproduces
---if one includes the phonons---the experimental spectrum. 
From this point of view, self-consistent calculations for excited states at the level of 
the 1p1h RPA are not appropriate.

\section{Acknowledgment}

We thank Sergey Kamerdzhiev for many useful comments. 
One of us (JS) thanks Stanislaw Dro\.zd\.z for many discussions and the Foundation for Polish 
Science for financial support through the 
\emph{Alexander von Humboldt Honorary Research Fellowship}.
The work was partly supported by the DFG and RFBR grants Nos.GZ:432RUS113/806/0-1
and 05-02-04005 and by the INTAS grand No.03-54-6545.

\end{document}